\documentclass[pra, twocolumn, floatfix,superscriptaddress]{revtex4-2}
\usepackage{graphicx}
\usepackage{epstopdf}
\usepackage{amsmath, amsfonts, amssymb, bm,color}
\def\ifa{Institute of Applied Physics, Moldova State University,
Academiei str. 5, MD-2028 Chi\c{s}in\u{a}u, Moldova}

\def\uniW{University of W\"{u}rzburg, Institute of Theoretical Physics and Astrophysics, Am Hubland, 97074 W\"{u}rzburg, Germany}
\begin{document}
\title{Phase-enhanced excitations in pumped collective nuclear systems } 
\author{Mihai A. Macovei }
\email{mihai.macovei@ifa.usm.md}
\affiliation{\ifa}

\author{Fabian Richter}
\affiliation{\uniW}

\author{Adriana P\'{a}lffy}
\email{adriana.palffy-buss@uni-wuerzburg.de}
\affiliation{\uniW}

\date{\today}
\begin{abstract}
The quantum dynamics of an externally driven ensemble of nuclear two-level systems embedded in a leaky broadband cavity is investigated theoretically.  In the considered scenario both the nuclear ensemble and the cavity mode are coherently pumped by two externally  applied x-ray electromagnetic fields. When the frequencies of the applied coherent fields are identical,  cross-correlations among the existing decay channels increase the 
nuclear excitation probabilities depending on the phase difference of the applied fields. Our results show that the excited state of the nuclear ensemble may exhibit sub- to super-Poissonian nuclear statistics, 
 demonstrating induced correlations during photon absorption or emission processes. The role of cross-correlations for the superradiant decay and the collective Lamb shift of the ensemble is also investigated. 
\end{abstract}
\maketitle

\section{Introduction}
While  quantum optics traditionally investigates how quantized optical light interacts with atoms or molecules, in the past decades this concept has extended also to x-ray or gamma-ray frequencies, and atomic nuclei as resonant matter system \cite{xqok,freqcomb}. Nuclear transitions have very narrow linewidths even at room temperature and can be very clean quantum optics systems, benefiting both quantum optics as well as metrology applications. For instance, challenging quantum optical experiments such as the measurement of the collective Lamb shift for single-photon superradiance \cite{expX11}, or the experimental proof of spontaneously generated coherences \cite{sgc} have been successfully performed using the 14.4 keV M\"ossbauer transition in $^{57}$Fe. Towards metrology applications, the very narrrow 12.4~keV transition of $^{45}$Sc has recently been excited with radiation from the X-Ray Free Electron Laser (XFEL) \cite{Shvydko2023}, while the low-lying $^{229}$Th isomer at only 8.4 eV promises a novel nuclear clock with potential to accurately test fundamental physics \cite{nuC}. 

Synchrotron radiation sources and the XFEL open new opportunities for resonant driving of ensembles of M\"ossbauer nuclei, which show interesting collective dynamics and allow for coherent control of single x-ray photons \cite{HeegNature2021}. M\"ossbauer nuclei such as $^{57}$Fe have been used as quantum optical few-level systems in either bulk or nanostructured targets. Thick targets are used in nuclear forward or Bragg scattering \cite{bible} at synchrotron radiation sources, and first experiments have been performed also at the XFEL \cite{Chumakov2018}. The sharp nuclear resonances have enabled high-fidelity, phase-coherent manipulation, which has facilitated  mechanical phase modulation \cite{Ruby1960,Perlow1978,Helisto1991,Shvydko1992,Heeg2017}, dynamical magnetic switching \cite{Shvydko1993,Shvydko1994,Shvydko1995,Shvydko1996}, and radio-frequency \cite{Lippmaa1995} as well as related time-domain techniques \cite{Shvydko1991,Hastings1991,vanB1992,HeegNature2021}.

\begin{figure}[t]
\includegraphics[width = 0.45\textwidth]{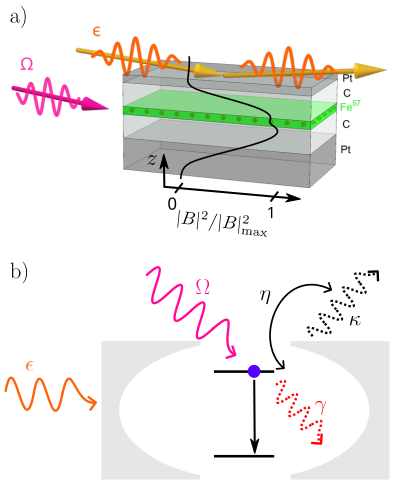}
\caption{\label{fig-0} 
(a) Thin-film cavity sketch. X-rays can couple in grazing incidence (orange field $\epsilon$), generating a standing wave in the cavity (black line), or front coupling (magenta field $\Omega$). The layer of M\"ossbauer nuclei is embedded in a cladding of high-$Z$ material (dark grey) and buffer low-$Z$ material (light grey). The latter does not interact with x rays and is  used for flexible positioning of the layer with M\"ossbauer nuclei in the cavity field. (b)  Simplified setup scheme. The quantum optical two-level systems are placed in a cavity and driven by two fields: the external field $\Omega$ and the cavity mode, driven in turn by the external field $\epsilon$. Cross-correlations couple the spontaneous decay $\gamma$ to the cavity decay channel $\kappa$.   }
\end{figure}

A different setup involves thin-film nanostructures with embedded layers of M\"ossbauer nuclei which are very interesting from theoretical quantum optics point of view due to their x-ray cavity or waveguide properties. In essence, these nanostructures consist of a thin film of a guiding 
low-atomic number $Z$ material, sandwiched  by two layers of a high-$Z$ material. One or several layers of M\"ossbauer nuclei can be embedded 
in the guiding layer. For resonant driving, the incident x-ray radiation is in resonance with the M\"ossbauer transition from the ground to the first 
nuclear excited state. The thin-film stack can be probed either in grazing incidence, with x-rays coupling evanescently and exciting guided modes, 
or in front or resonant beam coupling, with the structure behaving as an x-ray waveguide, as illustrated in Fig.~\ref{fig-0}(a). In grazing incidence geometry, the layered thin-film nanostructure can be modelled as a leaky cavity \cite{keJorg}. In this system, several interesting x-ray quantum optical phenomena based on strong collective effects have been investigated experimentally, for instance, superradiance and collective Lamb shift \cite{expX11}, electromagnetically induced transparency \cite{expX22}, spontaneously generated coherences \cite{sgc}, Rabi oscillations between two nuclear ensembles \cite{expX4}, or the strong coupling regime for x-rays \cite{Ralf2016}. The waveguide geometry with x-rays coupling from the side of the nanostructure has just recently been theoretically addressed \cite{Frontcoupling_theo} and experimentally demonstrated \cite{Frontcoupling_exp}. 

The new possibility to drive  the ensemble of identical M\"ossbauer nuclei with two beams using grazing incidence and front coupling geometries simultaneously raises interesting fundamental questions. Following pumping, emission into the broadband cavity mode may compete with decay channels outside this mode leading to cross-correlations among these decay 
paths; see for instance Refs.~\cite{PhysRevLett.70.2269,PhysRevLett.70.2273} for related effects. Combined with external pumping schemes, this can lead to  additional control tools of the nuclear quantum dynamics and to 
Kerr-like nonlinearities in the x-ray regime. Motivated by this, in this work we design and investigate theoretically a quantum optical toy model for an ensemble of identical two-level systems placed in a leaky broad-band cavity interacting with both cavity mode and an external driving field. We base our approach on the cavity QED formalism adapted for thin-film x-ray cavities \cite{keJorg}. In our generic model, an ensemble of two-level nuclei is embedded in a broad-band, leaky cavity, which is externally driven, similar to the grazing incidence setup. In addition, the ensemble is also externally driven by a second field,  resembling the x-ray thin-film in front coupling geometry, as illustrated in Fig.~\ref{fig-0}(a). In the bad cavity limit, the broadband cavity mode induces dipole-dipole interactions among the two-level emitters which subsequently modify the nuclear excitation's quantum dynamics 
and their second-order correlations in the steady-state. 

Our numerical results for generic parameters show that if the frequency of the cavity driving field is equal to the one of the external field driving the nuclei, the cross-correlations among the decaying channels lead to higher phase-dependent excitation probabilities as well as larger frequency shifts, even at the cavity resonance. Once the two frequencies are considerably different, the cross-correlation effects disappear. We focus on the regime of driving fields with the same frequency and   investigate the effect of cross-correlations on the superradiant decay rates and collective Lamb shifts, emphasizing the role of phase relation between the two driving fields on the steady-state nuclear excitation dynamics and its second-order correlation function. 

We find that, generally speaking, the number of excitations increases depending on the interference effects among the two applied external fields and cross-correlations within the decay channels. Enhanced excitations may induce Kerr-like nonlinear phenomena which can be probed if the excitation's normalized second-order correlation function deviates from  unity. In order to observe these effects, one would require at least two excitations in the nuclear ensemble. While at present synchrotron radiation pulses have typically at most just one resonant photon per pulse, XFEL excitation can provide several excitations per pulse \cite{Chumakov2018}. Thus, experiments on thin-film nanostructures with embedded M\"ossbauer nuclei at the XFEL can open a new platform for x-ray non-linear phenomena in future experiments.

This paper is organized as follows. In Sec.~\ref{theo} we introduce the analytical approach for the cases of equal and different driving field frequencies. Section~\ref{RD} presents and discusses numerical results. The article concludes with final remarks in Sec.~\ref{sum}.

\section{Theoretical framework \label{theo}}
At present, several theoretical approaches can successfully model the nuclear quantum dynamics in thin-film cavities. The Parratt formalism \cite{Parrat} or a self-consistent multiple scattering formalism using layer matrices \cite{rohlsbergerNuclearCondensedMatter2004} are semi-classical approaches with high predictive power but less quantum optical insight. 
The cavity QED formalism developed in Refs.~\cite{keJorg,HeegPRA2015}  requires fit para\-meters as input for quantitative predictions, but on the other hand naturally gives access to quantum optical quantities in the modeling. Finally, a more recently developed method based on the classical electromagnetic Green's function formalism to account for the interaction between embedded layers of M\"ossbauer nuclei has been very successful to model the nuclear excitation \cite{Green_fct_Xiangjin,Green_fct_Dominik}. The Green's function formalism has been used also in the context of front coupling geometry and waveguide propagation \cite{Frontcoupling_theo}. 

For our present study, an insightful description capable to address cross-correlations in an intuitive way is paramount. We therefore choose the cavity QED formalism and introduce in the following our theoretical quantum optics model. The system under investigation consists of an ensemble of $N$ externally pumped nuclei embedded in a thin-film broadband cavity with nm-thick layers of alternating high-$Z$ and low-$Z$ materials as illustrated in Fig.~\ref{fig-0}(a). Each nucleus is considered to be a two-level system with the excited (ground) state $|e\rangle$ ($|g\rangle$). The first coherent x-ray field with Rabi frequency $\Omega$ is pumping the nuclear solid-state sample in front coupling, similarly to the experimental setup described in Ref.~\cite{Frontcoupling_exp}, whereas the second one illuminates the cavity in grazing incidence, coupling evanescently and driving the cavity mode, see Fig.~\ref{fig-0}(b). For the most general 
case, we denote the  frequencies of the coherent driving fields as $\omega_{x1}$ and $\omega_{x2}$, and the corresponding field phases by $\phi_{1}$ and $\phi_{2}$, respectively. 

The Hamiltonian describing the pumped nuclear system in  the interaction picture, where the nuclear subsystem is rotating with angular frequency $\omega_{x1}$ and the cavity subsystem with $\omega_{x2}$, respectively,  can be written as
\begin{eqnarray}
H &=& \hbar \Delta_{c}a^{\dagger}a + \hbar\Delta_{n}S_{z} 
+ \hbar \Omega \bigl(S^{+}e^{-i\phi_{1}}
+ S^{-}e^{i\phi_{1}} \bigr) \nonumber \\
&+&\hbar\epsilon \bigl(a^{\dagger}e^{-i\phi_{2}} + ae^{i\phi_{2}} \bigr) + \hbar g_c \bigl(S^{+}ae^{i\Delta\omega t} \nonumber \\
&+& a^{\dagger}S^{-}e^{-i\Delta\omega t} \bigr), \label{nHm}
\end{eqnarray}
where $\hbar$ is the reduced Planck constant,  $\Delta_{c}=\omega_{c}-\omega_{x2}$ stands for the detuning between the cavity 
frequency $\omega_c$ and the field driving the cavity mode, and $\Delta_{n}=\omega_{n}-\omega_{x1}$ for the one between the 
nucleus transition frequency $\omega_{n}$ and the front coupling driving field frequency, respectively. The first term in the expression 
above stands for the free energy of the cavity, with $a$ ($a^{\dagger}$) the annihilation (creation) operators describing the cavity 
field, satisfying the standard bosonic commutation relations, i.e., 
$[a, a^{\dagger}] = 1$, and $[a,a]$ = $[a^{\dagger},a^{\dagger}]
= 0$ \cite{walls}. We introduced the collective nuclear spin operators 
\begin{eqnarray*}
S^{+} = \sum^{N}_{j=1}S^{+}_{j}=\sum^{N}_{j=1}|e_{j}\rangle\langle g_{j}|\,,
\end{eqnarray*}
and
\begin{eqnarray*}
S^{-} = \sum^{N}_{j=1}S^{-}_{j}=\sum^{N}_{j=1}|g_{j}\rangle\langle e_{j}|\, ,
\end{eqnarray*}
with $j$ indexing the two-level nuclei in the ensemble. The nuclear spin operators obey the  commutation relations for SU(2) algebra, namely, $[S^{+}_{j},S^{-}_{l}] =2S_{zj}\delta_{jl}$ and $[S_{zj},S^{\pm}_{l}]=\pm S^{\pm}_{l}\delta_{lj}$, with $S_{zj} = (|e_{j}\rangle\langle e_{j}|-|g_{j}\rangle\langle g_{j}|)/2$, 
or, similarly, for the collective operators: $[S^{+},S^{-}] =2S_{z}$ and $[S_{z},S^{\pm}]=\pm S^{\pm}$, with $S_{z} = \sum^{N}_{j=1}S_{zj}$ being 
the collective bare-state inversion operator. Hence, the second term of (\ref{nHm}) describes the free energy of nuclear subsystem. 

The nuclear transitions are driven directly by the external field with real Rabi frequency $\Omega$. This process is being described by the third term in the Hamiltonian \eqref{nHm}, whereas the fourth term accounts for the second external field characterized by the coupling strength to the cavity mode $\epsilon$. In addition, the fifth component characterizes the interaction between the broadband cavity electromagnetic field  and the nuclear ensemble with coupling constant $g_{c}$, while $\Delta\omega=\omega_{x1}-\omega_{x2}$. 

The Hamiltonian \eqref{nHm} governs the coherent dynamics of the system. 
The cavity and nuclear decay channels can be taken into account in the master equation approach, which includes decay terms \cite{reww,walls,gsag,carma,cross_rez}. 
From now on we focus on the more interesting case when the two external driving fields have the same frequencies. Hence, the master equation 
with corresponding damping terms included is given by
\begin{eqnarray}
\frac{d}{dt}\rho(t) &+& \frac{i}{\hbar}[H,\rho] = - \frac{\gamma}{2}\sum^{N}_{j=1}[S^{+}_{j},S^{-}_{j}\rho]- \frac{\kappa}{2}[a^{\dagger},a\rho] 
\nonumber \\
&-&\frac{\eta\sqrt{\kappa \gamma}}{2}[S^{+},a\rho] - \frac{\eta\sqrt{\kappa \gamma}}{2}[a^{\dagger},S^{-}\rho] \nonumber \\
&+& H.c.. \label{Meq}
\end{eqnarray}
The coherent evolution of the examined system is described by the second term on the left-hand side of the master equation (\ref{Meq}), where the Hamiltonian $H$ is given by (\ref{nHm}) with $\omega_{x1} =\omega_{x2}$. 
The damping effects of the involved nuclear and photon subsystems are characterized by the right-hand side of this equation. Particularly, the first two terms on the right-hand side describe the spontaneous and cavity decay processes with $\gamma$ and $\kappa$ being the corresponding nucleus or photon cavity damping rates, respectively. The last two terms from the master equation (\ref{Meq}) account for cross-correlation effects where a cavity photon is absorbed by a nucleus followed by its spontaneous emission in other modes,  or the nucleus spontaneously emits a photon, which is 
re-scattered into the broadband cavity mode and then leaks out through the cavity mirrors. The parameter $\eta$, with $0 \le \eta \le 1$, is unity for maximal cross-correlation effects and subunitary otherwise, while $\eta=0$ means no cross-correlation effects. The exact value of the cross-correlation parameter is not pertained in the model; however, a comparison with independent results obtained from the Green's function formalism can shed light on its order of magnitude, as discussed later on in Sec.~\ref{RD}.

In the following, we proceed to investigate the system's quantum dynamics in the weak excitation regime, when the nuclear excitations in the sample are smaller than the number of nuclei itself. We redefine the operators $S^{+}e^{-i\phi_{1}} \to S^{+}$ and 
$a^{\dagger}e^{-i\phi_{2}} \to a^{\dagger}$, and denote the new coupling strength between nucleus and cavity mode photons 
by $\bar g_c = g_ce^{i(\phi_{1}-\phi_{2})}$. Then using the unitary transformation
\begin{eqnarray}
\bar a^{\dagger} &=& D^{\dagger}(\alpha)a^{\dagger}D(\alpha)= a^{\dagger} + \alpha^{\ast}, \nonumber \\
\bar a &=& D^{\dagger}(\alpha)aD(\alpha)= a + \alpha, \label{disp}
\end{eqnarray}
where the displacement operator is given by $D(\alpha)=\exp{[\alpha a^{\dagger}-\alpha^{\ast}a]}$ with 
$\alpha=\epsilon/(\Delta_{c}-i\kappa/2)$, one arrives at the following master equation
\begin{eqnarray}
\frac{d}{dt}\rho(t) &+& \frac{i}{\hbar}[\bar H,\rho] = - \frac{\gamma}{2}\sum^{N}_{j=1}[S^{+}_{j},S^{-}_{j}\rho] - 
\frac{\kappa}{2}[\bar a^{\dagger},\bar a\rho] \nonumber \\
&-&\frac{\bar \gamma}{2}[S^{+},\bar a\rho] - \frac{\bar \gamma^{\ast}}{2}[\bar a^{\dagger},S^{-}\rho] + H.c..
\label{MeqT}
\end{eqnarray}
Here, $\bar \gamma = \eta\sqrt{\gamma \kappa}e^{i(\phi_{1}-\phi_{2})}$, 
and the Hamiltonian becomes
$\bar H = \bar H_{0}$+$\bar V$. The unperturbed Hamiltonian for nuclear ensemble and cavity field is given by
\begin{eqnarray}
\bar H_{0}=\hbar \Delta_{c}\bar a^{\dagger}\bar a + \hbar\Delta_{n}S_{z} + \hbar G S^{+} + \hbar G^{\ast}S^{-},
\label{H0T}
\end{eqnarray}
with $G=\Omega - \bar \epsilon$ and $\bar \epsilon= \epsilon(\bar g_c 
-i\bar \gamma/2)/(\Delta_{c}-i\kappa/2)$.
The interaction between the nuclear ensemble and the cavity mode is described by the interaction Hamiltonian
\begin{eqnarray}
\bar V = \hbar \bar g_c S^{+}\bar a + \hbar \bar g_c^{\ast} S^{-}\bar a^{\dagger}.
\label{VT}
\end{eqnarray}
In the bad cavity limit, as it is the case here, one can eliminate 
the field operators, i.e. $\kappa > \{\sqrt{N}g_c,N\gamma\}$. 
From the Heisenberg equations of motion for the cavity field variables 
one obtains the expressions
\begin{eqnarray*}
\bar a^{\dagger}=-\frac{i(\bar g_c + i\bar \gamma/2)}{i\Delta_{c}-\kappa/2}S^{+}, 
\end{eqnarray*}
\begin{eqnarray}
\bar a =-\frac{i(\bar g_c^{\ast} - i\bar \gamma^{\ast}/2)}{i\Delta_{c}+\kappa/2}S^{-}, \label{apm}
\end{eqnarray}
resulting in the following master equation describing the nuclear 
system alone
\begin{eqnarray}
\frac{d}{dt}\rho(t) &+& i\bigl[\Delta_{n}S_{z} + GS^{+} + G^{\ast}S^{-} + \bar\delta S^{+}S^{-},\rho \bigr] \nonumber \\
&=&-\frac{\gamma}{2}\sum^{N}_{j=1}[S^{+}_{j},S^{-}_{j}\rho] - \frac{\gamma_{c}}{2}[S^{+},S^{-}\rho ] + H.c..\nonumber \\
\label{MeqqT}
\end{eqnarray}
Here, $\bar \delta = - \delta_{c}$ with 
\begin{eqnarray}
\delta_{c}=\frac{\Delta_{c}(g_c^{2} - \kappa\gamma\eta^{2}/4)
+ \eta g_c\kappa\sqrt{\kappa\gamma}/2}{(\kappa/2)^{2}+\Delta^{2}_{c}}, \label{dc}
\end{eqnarray}
being the dipole-dipole interaction among the nuclei due to the broadband vacuum cavity mode. When multiplied with the number of nuclei $N$, $\delta_c$ becomes the collective Lamb shift associated with the superradiant speed-up of the collective nuclear decay \cite{expX11}. Furthermore, the cavity induces collective interactions among nuclei, while their contribution to the collective spontaneous decay rate becomes 
\begin{eqnarray}
\gamma_{c}=\frac{\kappa(g_c^{2}-\kappa\gamma\eta^{2}/4) - 2\eta g_c\Delta_{c}\sqrt{\kappa\gamma}}
{(\kappa/2)^{2}+\Delta^{2}_{c}}. \label{gmc}
\end{eqnarray}
If one has a single nucleus, then for $\eta=0$, $\gamma_{s}=\gamma + \gamma_{c}$ turns into the usual Purcell effect \cite{purcell}, namely,
\begin{eqnarray}
\gamma_{s}=\gamma + \frac{\kappa g_c^{2}}{(\kappa/2)^{2}+\Delta^{2}_{c}}. \label{get0}
\end{eqnarray}
If $\eta=1$, the decay rate turns into
\begin{eqnarray}
\gamma_{s}=\frac{\bigl (g_c\sqrt{\kappa} - \Delta_{c}\sqrt{\gamma}
\bigr)^{2}}{(\kappa/2)^{2}+\Delta^{2}_{c}}, \label{get1}
\end{eqnarray}
which is zero for $\Delta_{c}=g_c\sqrt{\kappa/\gamma}$. Thus, due to the cross-correlations among the spontaneous and cavity decay paths, the decay rate of an independent emitter, $\gamma_{s}$, can decrease or increase   depending on the chosen parameters. We emphasize that neither $\gamma_s$ nor $\delta_c$ explicitly depend on the external field $\Omega$ that directly drives the nuclear ensemble. The collective decay of the ensemble due to cavity mediated interaction between nuclei is described later.

The cavity mediated dipole-dipole interaction among the emitters, $\delta_{c}$, and therefore the collective Lamb shift also change accordingly. Particularly, $\delta_c$ vanishes for $\Delta_{c}=\eta=0$, while  for $\Delta_{c}=0$ and $\eta \not=0$ it yields 
\begin{eqnarray}
\delta_{c}=2\eta g_c\sqrt{\gamma/\kappa}. \label{dc0}
\end{eqnarray}
This result is consistent with the multimode approach developed in Ref.~\cite{fmJE} and also to experimental observations  of the collective Lamb shift \cite{expX11}. A  comparison with  results of the Green's function formalism \cite{Green_fct_Xiangjin,Green_fct_Dominik} will be presented for a numerical example in the next Section.

The master equation \eqref{MeqqT} can be solved analytically in steady-state using the approach developed in Refs.~\cite{kilin,puri} if the first term on the right-hand side is negligible compared to the second one. Here we take a different approach and keep both terms, restricting ourselves however to the low-excitation regime.  In order to investigate the collective Lamb shift, the collective decay rate and the Kerr-like nonlinearities for the realistic nuclear ensemble, we apply the Holstein-Primakoff transformation \cite{hprim}, namely,
\begin{eqnarray}
S^{+} &=& \sqrt{N}b^{\dagger}\sqrt{1-b^{\dagger}b/N}, \nonumber \\
S^{-} &=& \sqrt{N}\sqrt{1-b^{\dagger}b/N}b, \nonumber \\
S_{z} &=& b^{\dagger}b-N/2, \label{HP}
\end{eqnarray}
which introduces the bosonic operators $b$ ($b^{\dagger}$) describing the annihilation (generation) of an excitation in the sample. The operators satisfy the  usual bosonic commutation relations $[b,b^{\dagger}]=1$, and $[b,b]=[b^{\dagger},b^{\dagger}]=0$. With these transformations, the master equation describing the quantum dynamics of the nuclear system becomes
\begin{eqnarray}
\frac{d}{dt}\rho(t) &=&- i\bigl[\bar \Delta b^{\dagger}b + \bar \Omega b^{\dagger} + \bar \Omega^{\ast}b - \tilde\Omega b^{\dagger^2}b - \tilde\Omega^{\ast}b^{\dagger}b^{2} \nonumber \\
&-& \bar \delta b^{\dagger 2}b^{2},\rho\bigr] - \frac{\Gamma}{2}\bigl([b^{\dagger},b\rho] + [\rho b^{\dagger},b]\bigr) 
\nonumber \\
&+&\frac{\Gamma_{0}}{2}\bigl( [b^{\dagger^{2}}b,b\rho] + [\rho b^{\dagger}, b^{\dagger}b^{2}] \bigr). \label{Meqb}
\end{eqnarray}
Here, $\bar \Delta=\Delta_{n} - N\delta_{c}$, $\bar \Omega = \sqrt{N}G$, $\bar \Omega^{\ast} =\sqrt{N}G^{\ast}$, 
$\tilde \Omega=G/(2\sqrt{N})$, $\tilde \Omega^{\ast}=G^{\ast}/(2\sqrt{N})$,  $\Gamma=\gamma + N\gamma_{c}$ and $\Gamma_{0}=\Gamma/N$, where one recognizes the cooperative Lamb shift $N\delta_c$ and the collective 
superradiant decay rate $N\gamma_{c}$. In order to obtain Eq.~(\ref{Meqb}), we have used that $S^{+}S^{-}=Nb^{\dagger}b - b^{\dagger 2}b^{2}$ in the dipole-dipole interaction term, while in other cases we have approximated $S^{+} \approx \sqrt{N}b^{\dagger}\bigl(1 - b^{\dagger}b/2N\bigr)$ and $S^{-} \approx \sqrt{N}\bigl(1-b^{\dagger}b/2N\bigr)b$, respectively, i.e. assuming the weak excitation regime, $b^{\dagger}b/N \ll 1$. Notice that 
a positive decay rate $\Gamma$, i.e. $\Gamma \ge 0$, will require $\eta \le 1/\sqrt{N}$.

Solving the master equation (\ref{Meqb}) is a challenging task. However, for the particular case we consider here, it can be solved analytically in the steady-state limit using the approach described in Refs.~\cite{walls,drum}. 
For this, we rewrite this equation in the form of a Fokker-Planck equation in the generalized $P$ representation, namely,
\begin{eqnarray}
\frac{\partial{P(\beta)}}{\partial{t}} &=& \frac{\partial}{\partial \beta_{1}}\biggl\{\bigl(\Gamma/2+i\bar \Delta\bigr)\beta_{1} 
+ i\bar\Omega - 2i\tilde\Omega \beta_{1}\beta_{2} -i\tilde\Omega^{\ast}\beta^{2}_{1}\nonumber \\
&-&\bigl(\Gamma_{0}/2 + 2i\bar\delta \bigr)\beta^{2}_{1}\beta_{2} 
+ \bigl(\Gamma_{0}/2 + i\bar\delta \bigr)\frac{\partial}{\partial \beta_{1}}\beta^{2}_{1} \nonumber \\
&+& i\tilde\Omega\frac{\partial}{\partial \beta_{1}}\beta_{1}\biggr\}P(\beta) \nonumber \\
&+& \frac{\partial}{\partial \beta_{2}}\biggl\{\bigl(\Gamma/2 - i\bar \Delta\bigr)\beta_{2} 
- i\bar\Omega^{\ast} + 2i\tilde\Omega^{\ast}\beta_{1}\beta_{2} + i\tilde\Omega\beta^{2}_{2}\nonumber \\
&-&\bigl(\Gamma_{0}/2 - 2i\bar\delta \bigr)\beta^{2}_{2}\beta_{1} 
+ \bigl(\Gamma_{0}/2 - i\bar\delta \bigr)\frac{\partial}{\partial \beta_{2}}\beta^{2}_{2}\nonumber \\
&-& i\tilde\Omega^{\ast}\frac{\partial}{\partial \beta_{2}}\beta_{2} \biggr\}P(\beta),
\label{FPE}
\end{eqnarray}
where $\beta \in \{\beta_{1},\beta_{2}\}$ are the phase-space coordinates 
of the nuclear subsystem, while $P(\beta)$ is a quasiprobability distribution over the phase space $\beta$, respectively. The analytical solution in steady-state exists only for the particular case of $\Gamma_{0}=0$. Since $\Gamma_{0}/\Gamma=1/N$, and $N$ can generally speaking be a large number, it is reasonable to neglect the terms containing $\Gamma_{0}$. 
The analytical  solution is then given by,
\begin{eqnarray}
P(\beta)&=&Z^{'-1}(\beta_{1}\beta_{2})^{-1-2N}\bigl(\tilde\Omega + \bar\delta\beta_{1}\bigr)^{\Theta}
\bigl(\tilde\Omega^{\ast} + \bar\delta\beta_{2}\bigr)^{\Theta^{\ast}}\nonumber \\ 
&\times& e^{2\beta_{1}\beta_{2}+\tilde\Omega^{\ast}\beta_{1}/\bar\delta + \tilde\Omega\beta_{2}/\bar\delta},
\label{ssl}
\end{eqnarray}
where $Z^{'}$ is a normalization constant, while $\Theta=1+2N - p + i\bar\Gamma$ and $p=|\tilde\Omega|^{2}/\bar\delta^{2}$,
$\bar\Gamma=\bigl(\Gamma/2+i(\bar\Delta+2\bar\delta) \bigr)/\bar\delta$ with $\bar\delta \not=0$, respectively. The special case $\bar\delta\approx 0$ needs to be considered separately and is discussed in  Appendix \ref{appA}.

Using the long-time solution (\ref{ssl}), one can obtain the following expression for the mean excitation number in the sample 
\begin{eqnarray}
\langle b^{\dagger}b\rangle &=&(2Z)^{-1}\sum_{n}\frac{(2p)^{n}}{n!}n|I_{n}|^{2}, \label{bpm} \\
Z &=& \sum_{n}\frac{(2p)^{n}}{n!}|I_{n}|^{2}, \nonumber 
\end{eqnarray}
or the nuclear excitation amplitude 
\begin{eqnarray}
\langle b\rangle = \frac{\tilde\Omega}{\bar\delta Z}\sum_{n}\frac{(2p)^{n}}{n!}I_{n+1}I^{\ast}_{n}, \label{ba} 
\end{eqnarray}
as well as the nuclear  second-order correlation function, respectively, 
\begin{eqnarray}
\langle b^{\dagger 2}b^{2}\rangle &=&(4Z)^{-1}\sum_{n}\frac{(2p)^{n}}{n!}n(n-1)|I_{n}|^{2}. \label{bbpm} 
\end{eqnarray}
The normalized second-order nuclear correlation function \cite{glbr} for nuclei residing in their excited state, is  then given by
\begin{eqnarray}
g^{(2)}_{b}(0)=\langle b^{\dagger 2}b^{2}\rangle/
\langle b^{\dagger}b\rangle^{2}. \label{g2n}
\end{eqnarray}
Here,
\begin{eqnarray*}
 I_{n} = \binom{\Theta}{2N-n}{}_{1}F_{1}(1 + \Theta, 1 + \Theta + n - 2N;p), 
\end{eqnarray*}
and
\begin{eqnarray*}
{}_{1}F_{1}(a,b;z) = \sum^{\infty}_{k=0}\frac{\Gamma(a+k)\Gamma(b)}{\Gamma(a)\Gamma(b+k)}\frac{z^{k}}{k!},
\end{eqnarray*}
is the generalized hypergeometric function, with $\Gamma(x)$ being the  
gamma function. 

The mean excitation number as well as the second-order correlation 
function depend on $p$, while $p \sim N|G|^{2}$, respectively. 
However, 
\begin{eqnarray}
|G|^{2}&=&\Omega^{2} + \frac{\epsilon^{2}\bigl(g^{2}_{c}+\eta^{2}\gamma\kappa/4\bigr)}{\Delta^{2}_{c}+\kappa^{2}/4} + 
\frac{2\epsilon\Omega}{\kappa^{2}+(2\Delta_{c})^{2}} \nonumber \\
&\times& \bigl(2(\kappa g_{c}-\eta\Delta_{c}\sqrt{\gamma\kappa})\sin{\Delta\phi} - (\kappa\eta\sqrt{\gamma\kappa}
+ 4g_{c}\Delta_{c}) \nonumber \\
&\times& \cos{\Delta\phi}\bigr), 
\label{ggm}
\end{eqnarray}
where $\Delta\phi=\phi_{1}-\phi_{2}$. Therefore, both applied fields $\Omega$ and $\epsilon$ contribute to the final steady-state 
dynamics, and so do the cross-correlation damping terms proportional 
to $\eta$. Actually, the excitation number may increase or diminish, depending on the interference effects among the two applied external 
fields or on cross-correlation decay channels.

If one neglects the nonlinear as well as the Kerr-like terms from the 
master equation (\ref{Meqb}), i.e.,  the terms proportional to 
$\{\bar\delta,\tilde\Omega,\tilde\Omega^{\ast}\}$, then the mean excitation number and the  second-order correlation function are given by $\langle b^{\dagger}b\rangle=
N|G|^{2}/[(\Gamma/2)^{2} + N|G|^{2}]$ and $\langle b^{\dagger 2}b^{2}\rangle$ = $N^{2}|G|^{4}/[(\Gamma/2)^{2} + N|G|^{2}]^{2}$,
respectively.  We note that in this case the normalized second-order nuclear excitation correlation function equals unity, 
$g^{(2)}(0)$=$\langle b^{\dagger 2}b^{2}\rangle/\langle b^{\dagger}b\rangle^{2}=1$. Thus, the deviation of the normalized 
second-order correlation function from unity is a signature of the nonlinear contribution to the established 
steady-state quantum dynamics, and indicates that the probability for at least two excitations in the nuclear ensemble is not negligible.

\section{Results and Discussion \label{RD}}
In this Section, based on Eqs.~(\ref{bpm}) and (\ref{g2n}),  we investigate in numerical simulations the pumped nuclear population quantum dynamics and  its correlations in the weak excitation regime. We start by calculating the mean number of nuclear excitations in the ensemble as a function of the cavity detuning $\Delta_c$ according to Eq.~(\ref{bpm}) for a set of generic parameters. The results are presented in Fig.~\ref{fig-1} for both cases with maximal $(\sqrt{N}\eta=1)$ and without cross-correlations $(\eta=0)$; for the latter case  we use the expressions from  Appendix \ref{appA}. At this stage, we  consider just one of the two driving fields present (and therefore no effect of phase differences $\Delta \phi$) to support later comparisons.  The maximum excitation occurs for the case of $\eta=0$ at $\Delta_{c}/\gamma = - (\omega_{n} - \omega_{c})/\gamma + \Delta_{n}/\gamma \approx - (\omega_{n} - \omega_{c})/\gamma  \approx -50$, regardless of which of the driving fields was considered in the calculation. At the maximum, the applied laser frequency  is close to resonance with the nuclear transition 
frequency $\omega_{n}$, i.e. $\Delta_{n}/\gamma \approx 0$ and the collective Lamb shift $N\delta_{c}$ is negligibly small.  
\begin{figure}[t]
\includegraphics[width =4cm]{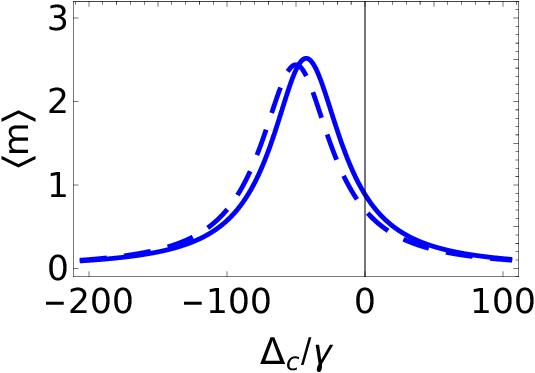}
\hspace{1mm}
\includegraphics[width =4.2cm]{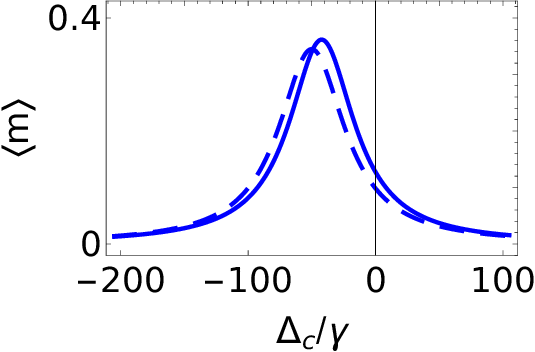}
\begin{picture}(0,0)
\put(-225,68){(a)}
\put(-25,68){(b)}
\end{picture}
\caption{\label{fig-1} 
The mean nuclear excitation number $\langle m\rangle \equiv \langle b^{\dagger}b\rangle$ as a function of  $\Delta_{c}/\gamma$ 
for (a) $\sqrt{N}\Omega/\gamma=50$, $\epsilon/\gamma=0$, and (b) $\sqrt{N}\Omega/\gamma=0$, $\epsilon/\gamma=3\times 10^{3}$. 
Other involved parameters are: $\kappa/\gamma=162 \times 10^{4}$, $\sqrt{N}g/\gamma = 5.02\times 10^{3}$, $(\omega_{n}-\omega_{c})/\gamma=50$. The dashed line is plotted for $\eta=0$ whereas the solid one for $\sqrt{N}\eta=1$.}
\end{figure}
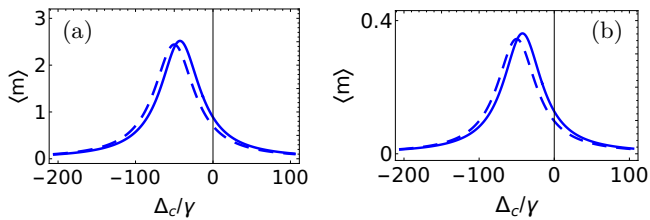
\begin{figure}[b]
\includegraphics[width = 7cm]{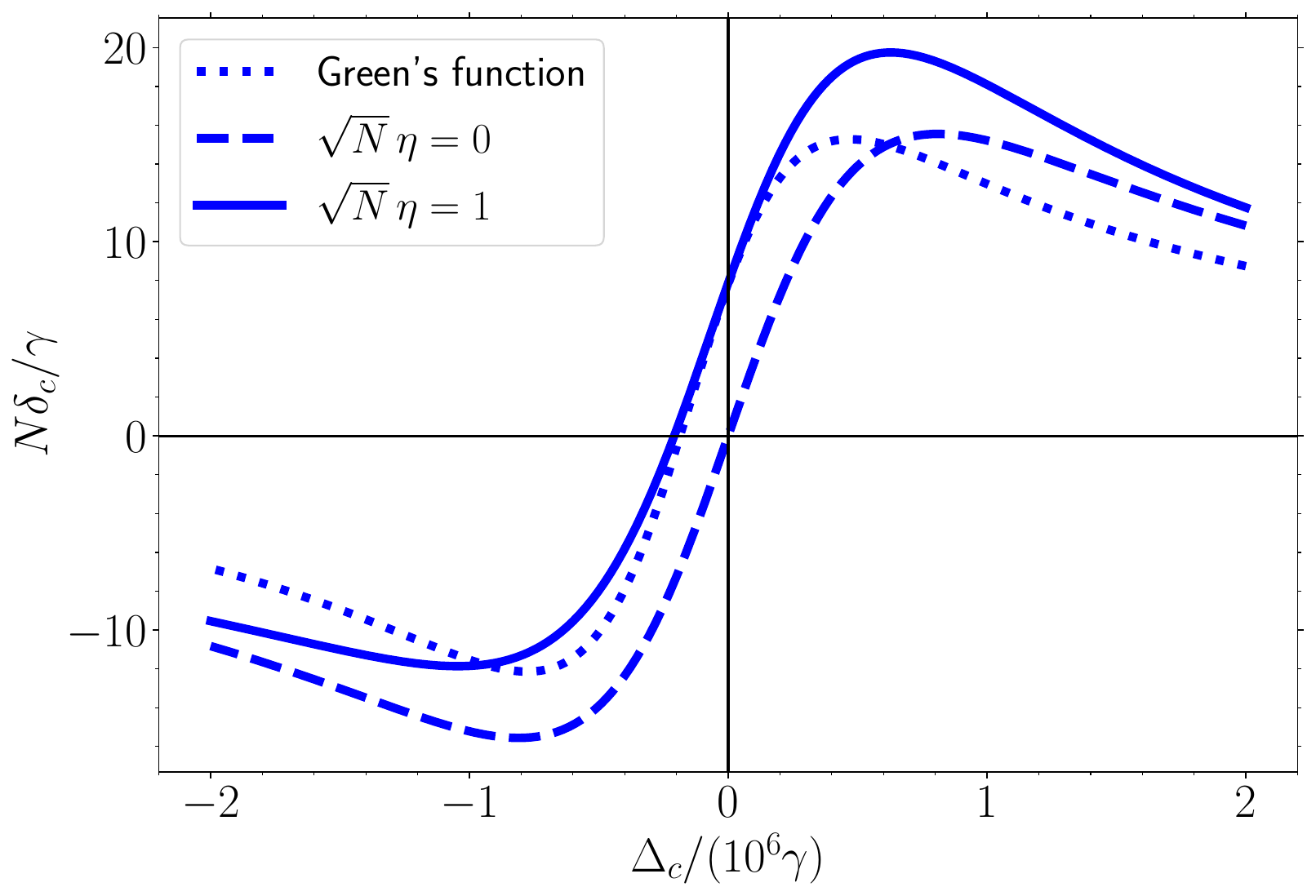}
\caption{\label{fig-2} 
The scaled collective Lamb shift $N\delta_{c}/\gamma$ based on Eq.~(\ref{dc}) as a function of $\Delta_{c}/\gamma$ for $\eta=0$ (dashed line) and $\sqrt{N}\eta=1$ (solid line). The parameters are the same as the ones used in Fig.~\ref{fig-1}. For comparison, the scaled collective Lamb shift calculated for the same parameters using the ab-initio Green's function formalism \cite{Green_fct_Xiangjin,Green_fct_Dominik} is presented by the dotted line. The thin-film cavity layer parameters used for the Green's function formalism were taken from Ref.~\cite{keJorg}. } 
\end{figure}

In comparison, when including the cross-correlations with $\sqrt{N}\eta=1$, the maximum for the nuclear excitation shifts to 
$\bar \Delta = \Delta_{n}-N\delta_{c}\approx 0$, leading 
to $\Delta_{n} \approx N\delta_{c}$. Apart from this shift, the cross-correlations between the decay paths  slightly enhance the excitation probability. In order to better understand the behavior of the collective Lamb shift, we investigate its dependence on the cavity detuning for both cases with and without cross-correlations in Fig.~\ref{fig-2} using the same set of generic parameters given for Fig.~\ref{fig-1}. In addition, for a comparison we include also the result of the Green's function formalism calculation following Refs.~\cite{Green_fct_Xiangjin,Green_fct_Dominik} for the same parameters. The Green's function formalism actually requires concrete thin-film layer parameters such as atomic numbers, layer thicknesses and densities for the calculation. The correspondence between the cavity model parameters used here (see caption of Fig.~\ref{fig-1}) and the thin-film layer structure is given by using the concrete numerical case considered in Ref.~\cite{keJorg} where both parameter sets are given. 

The presence of cross-correlations introduces a shift of the collective Lamb shift for a larger range of cavity detuning centered on $\Delta_c=0$. Around the resonance frequency, in the interval 
$[-10^6\gamma, \, 0.4\times10^6\gamma]$ the Green's function formalism agrees very well with the collective Lamb shift calculated for the case $\sqrt{N}\eta=1$, thus providing an independent confirmation for the role of cross-correlations and their magnitude. At very large detunings, the ab initio Green's function result deviates from both other curves. We conclude that the effects of  cross-correlations on the collective Lamb shift are visible, emphasizing the need to  account for cross-correlations in the cavity QED model using  $\sqrt{N}\eta=1$.

\begin{figure}[t]
\includegraphics[width = 4cm]{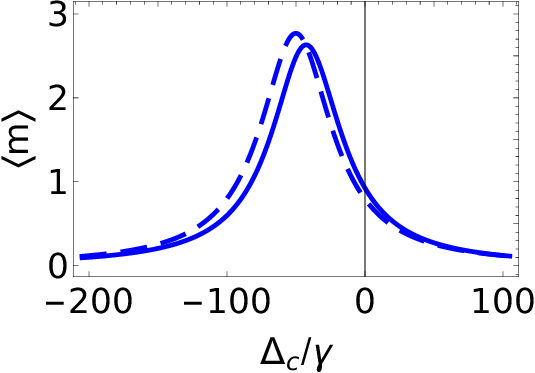}
\hspace{1mm}
\includegraphics[width = 4.2cm]{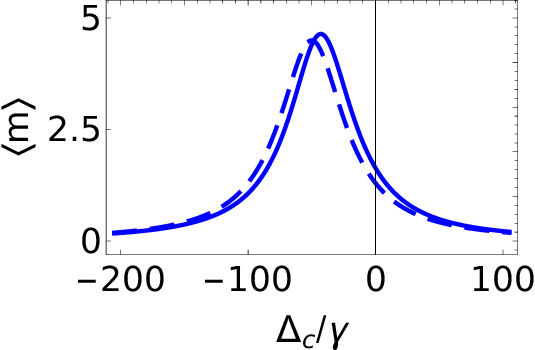}
\begin{picture}(0,0)
\put(-225,68){(a)}
\put(-25,68){(b)}
\end{picture}
\caption{\label{fig-3}
The mean nuclear excitation number $\langle m\rangle \equiv \langle b^{\dagger}b\rangle$ as a function of $\Delta_{c}/\gamma$ when both fields are applied simultaneously, for  (a)  $\Delta \phi = 0$, (b) $\Delta  \phi = \pi/2$.  We use $\sqrt{N}\Omega/\gamma=50$ and $\epsilon/\gamma=3\times 10^{3}$, while all other involved parameters are 
as in Fig.~\ref{fig-1}. Also here, the dashed line is plotted for $\eta=0$ whereas the solid one for $\sqrt{N}\eta=1$.}
\end{figure}
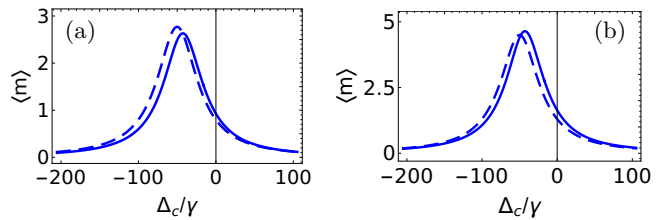

We now turn to the case when both fields are applied simultaneously to the nuclear ensemble. Numerical results for the mean nuclear excitation number are presented in Fig.~\ref{fig-3} for the dimensionless field parameters $\sqrt{N}\Omega/\gamma=50$ and $\epsilon/\gamma=3\times 10^{3}$.  The magnitude of the excitation peak is higher with respect to the case 
when only one field drives the sample. It can be even further enhanced depending on the phase difference of the externally applied 
coherent x-ray laser sources, cf. Figs.~\ref{fig-1} and ~\ref{fig-3}. In all cases, the peaks are broadened and proportional to 
$N$, as expected because of the superradiance phenomenon.

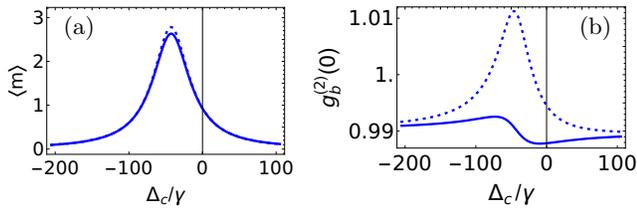
\begin{figure}[t]
\includegraphics[width = 3.8cm]{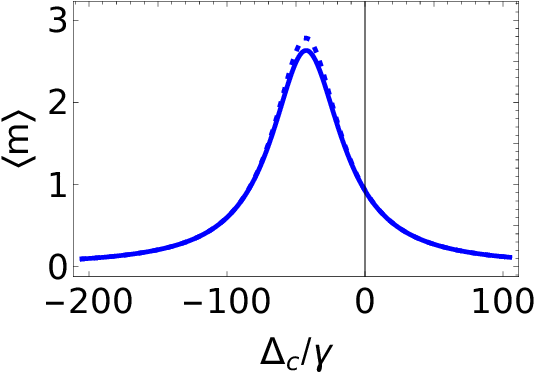}
\hspace{0.5mm}
\includegraphics[width = 4.3cm]{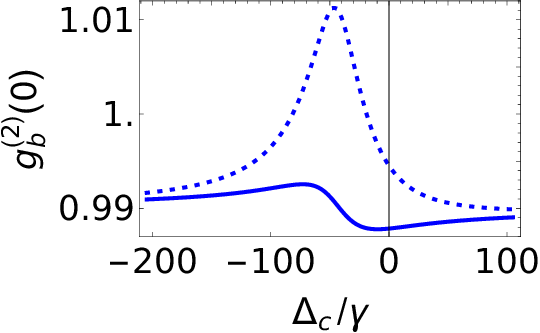}
\begin{picture}(0,0)
\put(-220,65){(a)}
\put(-25,65){(b)}
\end{picture}
\caption{\label{fig-4} 
(a) The mean nuclei excitation number $\langle m\rangle$ and (b) the nuclear second-order corelation function 
$g^{(2)}_{b}(0)$ as a function of  $\Delta_{c}/\gamma$. The solid lines are plotted  using Eqs.~(\ref{bpm}), (\ref{bbpm}) and (\ref{g2n}) 
with the approximation $\Gamma_{0}=0$,  whereas the dotted curves are obtained using the numerical solution of  
Eq.~(\ref{Meqb}), with $\Gamma_{0}=\Gamma/N$. Other parameters are the same as in Fig.~\ref{fig-3}(a) 
with $\sqrt{N}\eta=1$.  }
\end{figure}

The results presented so far considered the analytical solution \eqref{bpm}, which neglects the nonlinear contribution represented by the terms multiplying $\Gamma_{0}$ in the master equation Eq.~(\ref{Meqb}). In the following, we  compare the analytical results presented so far for the case of cross-correlations $\sqrt{N}\eta=1$ with those obtained from the numerical solution of Eq.~(\ref{Meqb}) without neglecting the $\Gamma_0$ terms. Our results are presented in Fig.~\ref{fig-4}, where the solid lines 
depict the steady-state behavior obtained by using Eq.~(\ref{bpm}), whereas the dotted curves show the numerical solution of Eq.~(\ref{Meqb}), with $\Gamma_{0}=\Gamma/N$. In addition to the mean nuclear excitation number, we also present the second order correlation function given by Exp.~\eqref{g2n}. Our results show that the mean nuclear excitation numbers are almost the same whether or not the terms containing $\Gamma_0$ are neglected, with small deviations around the maximum at  
$\Delta_n \approx N\delta_{c}$. This confirms that the nuclear excitation is  adequately described by the approximate analytical solution in Eq.~(\ref{bpm}). However, this is not the
case for the second-order correlation function $g^{(2)}_{b}(0)$, where the solid and dotted curves differ appreciably.
This is because the nonlinear terms multiplying $\Gamma_{0}$ do contribute to $g^{(2)}_{b}(0)$ which is a nonlinear
function by itself. The numerical exact second-order correlation function exhibits both sub-Poissonian ($g^{(2)}_{b}(0)<1$) as well as super-Poissonian ($g^{(2)}_{b}(0) >1$) nuclear statistics 
depending on the value of the cavity detuning. The effect occurs due to nonlinearity induced by 
the cavity mediated dipole-dipole interaction among the nuclei and is absent for the appoximate analytical solution which remains sub-Poissonian for all detunings. 

Thus, the cooperative interaction among nuclei, mediated by a strongly leaking broadband cavity mode, is responsible for 
the decay widths proportional to the number of emitters $N$ as well as for sub- to super-Poissonian nuclear statistics, while 
the cross-correlations among the existing decay channels induces frequency shifts dependent on the ensemble sizes even at 
$\Delta_{c}=0$. 

\section{Conclusions \label{sum}}
We have investigated the collective quantum dynamics of an externally pumped  ensemble of nuclear two-level systems embedded in a 
broadband leaking cavity, whose mode is also pumped  with a coherent electromagnetic field. This scenario is related to a thin-film x-ray cavity with an embedded layer of resonant M\"ossbauer nuclei under simultaneous grazing incidence and front coupling illumination. The cross-correlations 
among the involved decay channels are relevant when the frequencies of the applied fields are equal or when only a single pumping external coherent field is considered. As a consequence, in the bad cavity limit, the ensemble's collective spontaneous decay rate and resulting frequency shifts modify accordingly  depending on the sample's parameters, leading to higher 
phase-dependent excitation peaks and nonzero values for the collective Lamb shift even at cavity resonance. This behaviour of the collective Lamb shift has been confirmed experimentally in x-ray thin-film cavities \cite{expX11,eit_exp,keJorg}. An independent comparison with an ab-initio calculation using the Green's function formalism confirms that for a large detuning range cross-correlations with $\sqrt{N}\eta=1$ should be included in the cavity QED quantum optical model used.
Finally, the second-order correlation function for excited state nuclei changes from super- to sub-Poissonian nuclear statistics, demonstrating correlations among the emitters during absorption or emission processes, respectively. The super-Poissonian behaviour is related to non-linear effect and Kerr-like non-linearities which should be rendered possible in the future by XFEL fields which promise multiple nuclear excitations per pulse \cite{Chumakov2018} and a new access to superradiance studies \cite{Xiangjin2026}.

\acknowledgments
We thank  P. Andrejić, W.-T. Liao and D. Lentrodt for useful discussions.  M. M. gratefully acknowledges the financial support 
of the Alexander von Humboldt Foundation, Germany, as well as from the Moldavian Ministry of Education and Research, grant No. 011205. 
This work was supported by the German Science Foundation
(Deutsche Forschungsgemeinschaft, DFG) in the framework
of the Cluster of Excellence on Complexity and Topology in Quantum Matter ct.qmat (EXC 2147, Project No.
390858490). A.P. gratefully acknowledges the Heisenberg
Program of the DFG (Project No. 435041839).

\appendix
\section{Weak Kerr-like effects \label{appA}}
We consider here separately the case of very small $\bar\delta$, which occurs if $\eta=0$, see Fig.~(\ref{fig-2}) around 
$\Delta_{c}/\gamma=0$. In this case, we can neglect the term $\bar\delta b^{\dagger2}b^{2}$ from Eq.~(\ref{Meqb}), which is equivalent with the omission of the terms $\pm i\bar\delta(\partial/\partial \beta_{1,2})\beta^{2}_{1,2}P(\beta)$ from the Fokker-Planck equation (\ref{FPE}). Then, the solution of the Fokker-Plank equation for $\Gamma_{0}=0$ is
\begin{eqnarray}
P=Z^{''-1}(\beta_{1}\beta_{2})^{-(1+2N)}e^{2\beta_{1}\beta_{2} + f\beta^{2}_{1} + f^{\ast}\beta^{2}_{2} 
- \tilde\Gamma\beta_{1} - \tilde\Gamma^{\ast}\beta_{2}}. \nonumber \\
\label{s0l}
\end{eqnarray}
Here, $\tilde \Gamma=(\bar\Delta - i\Gamma/2)/\tilde\Omega$, $\tilde \Gamma^{\ast}=(\bar\Delta + i\Gamma/2)/\tilde
\Omega^{\ast}$,
$f=\tilde\Omega^{\ast}/(2\tilde\Omega)$ and $f^{\ast}=\tilde\Omega/(2\tilde\Omega^{\ast})$.

The normalization constant $Z^{''}$ is obtained from the condition 
\begin{eqnarray}
Z^{''}&=&\int d\beta_{1}\int d\beta_{2}(\beta_{1}\beta_{2})^{-(1+2N)}e^{2\beta_{1}\beta_{2}} \nonumber \\
&\times&e^{f\beta^{2}_{1} + f^{\ast}\beta^{2}_{2}-\tilde\Gamma\beta_{1} - \tilde\Gamma^{\ast}\beta_{2}}, 
\label{zzc}
\end{eqnarray}
or
\begin{eqnarray}
Z^{''}=\sum^{\infty}_{n=0}\frac{2^{n}}{n!}\biggl\vert\int d\beta\beta^{n-(1+2N)}
e^{f\beta^{2}-\tilde\Gamma\beta}\biggr\vert^{2}.
\label{zzz}
\end{eqnarray}
The integral entering  (\ref{zzz}) is evaluated using the residue theorem, namely,
\begin{eqnarray}
&{}&\int d\beta\beta^{n-(1+2N)}e^{f\beta^{2}-\tilde\Gamma\beta}=\sum_{k_{1},k_{2}}\frac{f^{k_{1}}(-\tilde\Gamma)^{k_{2}}}
{k_{1}!k_{2}!} \nonumber \\
&\times&\oint_{C} \frac{d\beta}{\beta^{1+2N-n-2k_{1}-k_{2}}}=2\pi i\sum_{k}\frac{f^{k}(-\tilde\Gamma)^{2N-n-2k}}{k!(2N-n-2k)!}
\nonumber \\
&=&i\pi 2^{2N-n}(-\tilde\Gamma)^{2N-n}(-\tilde\Gamma^{2}/f)^{(1+n-2N)/2} \nonumber \\
&\times&\frac{U[(1-2N+n)/2,3/2;-\tilde\Gamma^{2}/(4f)]}{(2N-n)!}, \nonumber
\end{eqnarray}
where $U$ is the corresponding hypergeometric function. Hence,
\begin{eqnarray}
Z^{''}=K\sum_{n}\frac{2^{-n}}{n!}|\tilde I_{n}|^{2} \equiv K\tilde Z,
\end{eqnarray}
where 
\begin{eqnarray*}
K = \pi^{2}2^{4N}|\tilde\Gamma|^{2}|f|^{-(1-2N)}, 
\end{eqnarray*}
and
\begin{eqnarray*}
\tilde Z = \sum_{n}\frac{2^{-n}}{n!}|\tilde I_{n}|^{2}, 
\end{eqnarray*}
whereas
\begin{eqnarray*}
\tilde I_{n}=\frac{(-f)^{-n/2}}{(2N-n)!}U[(n+1-2N)/2,3/2;-\tilde\Gamma^{2}/(4f)].
\end{eqnarray*}
Taking into account that 
\begin{eqnarray*}
\langle b^{\dagger l}b^{j}\rangle &=&Z^{''-1}\int d\beta_{1}\int d\beta_{2}\beta^{l}_{1}\beta^{j}_{2}
(\beta_{1}\beta_{2})^{-(1+2N)}e^{2\beta_{1}\beta_{2}} \nonumber \\
&\times&e^{f\beta^{2}_{1}+f^{\ast}\beta^{2}_{2}-\tilde\Gamma\beta_{1} - \tilde\Gamma^{\ast}\beta_{2}},
\end{eqnarray*}
${\rm \{l,j \in 0,1,2, \cdots \}}$, one can obtain the following expressions for the mean excitation number, 
\begin{eqnarray}
\langle b^{\dagger}b\rangle = \frac{1}{2\tilde Z}\sum_{n}\frac{2^{-n}}{n!}n|\tilde I_{n}|^{2}. 
\label{b0pm}
\end{eqnarray}
The solution (\ref{b0pm}) was used in plotting the dashed curves for $\eta=0$ in Figs.~\ref{fig-1} and \ref{fig-3}, which also coincide  with the corresponding ones obtained from the numerical solution of the master equation (\ref{Meqb}) for the same parameters.

\bibliographystyle{apsrev4-2} 
\bibliography{literatur.bib}

\end{document}